\documentclass[aps]{revtex4}
\usepackage[english]{babel}
\usepackage{multirow}
\usepackage[mathcal]{eucal}
\usepackage{rotating}
\usepackage{graphicx}
\usepackage{dcolumn}
\usepackage{bm}
\def\beq{\begin{equation}}
\def\eeq{\end{equation}}
\def\beqn{ \begin{eqnarray} }
\def\eeqn{ \end{eqnarray} }
\def\s1s2{{ \boldsymbol{\sigma}(1) \cdot \boldsymbol{\sigma}(2) }}
\def\t1t2{{ \boldsymbol{\tau}(1) \cdot \boldsymbol{\tau}(2)  }}
\newcommand{\vchi}{v_{\chi}}
\newcommand{\vz}{v_0}
\newcommand{\vmin}{v_{\rm min}}
\newcommand{\mchi}{M_{\chi}}
\newcommand{\muchia}{\mu_{\chi A}}
\newcommand{\ma}{M_{A}}
\newcommand{\er}{E_R}
\newcommand{\etr}{E_{\rm th}}
\newcommand{\fhelm}{F_{\rm Helm}}
\newcommand{\gnp}{g_{np}}

\newcommand{\oxy}{$^{16}$O}
\newcommand{\ca}{$^{40}$Ca}
\newcommand{\pb}{$^{208}$Pb}
\newcommand{\ar}{$^{40}$Ar}
\newcommand{\ger}{$^{72}$Ge}
\newcommand{\xe}{$^{136}$Xe}

\newcommand{\half}{\frac{1}{2}}

\newcommand{\br}{{\bf r}}
\newcommand{\bq}{{\bf q}}

\begin{document} 

\title{Nuclear proton and neutron distributions in the detection of
  weak interacting massive particles }
\author{G. Co'$^a$, V. De Donno$^a$, M. Anguiano$^b$, A. M. Lallena$^b$}
\affiliation{$^a$ Dipartimento di Matematica e Fisica 
``E. De Giorgi'', Universit\`a del Salento, 
Lecce, Italy,  and  
INFN, Sezione di Lecce, Via Arnesano, I-73100 Lecce, Italy \\
$^b$Departamento de F\'\i sica At\'omica, Molecular y
  Nuclear, Universidad de Granada, E-18071 Granada, Spain}
\date{\today}

\begin{abstract} In the evaluation of weak interacting massive particles
  (WIMPs) detection rates, the WIMP-nucleus cross section is commonly
  described by using form factors extracted from charge
  distributions. In this work, we use different proton and neutron
  distributions taken from Hartree-Fock calculations. We study the
  effects of this choice on the total detection rates for six nuclei
  with different neutron excess, and taken from different regions of
  the nuclear chart. The use of different distributions for protons
  and neutrons becomes more important if isospin-dependent
  WIMP-nucleon interactions are considered.  The need of distinct
  descriptions of proton and neutron densities reduces with the
  lowering of the detection energy thresholds.
 \end{abstract} 
\maketitle
\section{Introduction}
\label{sec:intro}

The elastic collision with a nucleus is the main mechanism used by
modern detectors to identify neutralinos, or more in general weakly
interacting massive particles (WIMPs)
\cite{ber10,aal11,ahm11,apr11,arm11}.  The cross section describing
this collision can be separated in a part related to the interaction
between the WIMP and the single nucleon, and another one describing
how the struck nucleon is inserted inside the nucleus \cite{eng92}.
While the former term is one of the unknown which we aim to study in
this type of investigations, the knowledge of the latter term is
solid, and well grounded on many years of nuclear structure
investigation.

The procedures commonly adopted substitute the nucleon distribution
inside the nucleus with the charge distribution which is
experimentally known with high spatial resolution because it is
extracted from the large body of accurate elastic electron scattering
data. In specific applications the Helm form factor \cite{hel56} is
usually adopted because of its simple analytic expression whose
parameters have been chosen to provide a good description of the
empirical charge distributions overall the nuclear chart, but for very
light nuclei. An improvement with respect to this approach has been
proposed in Ref. \cite{dud07} where it was suggested to consider
directly the measured charge distributions from the compilation of
Ref. \cite{dej87}.

The use of the experimental charge distributions in the calculations
of the WIMP-nucleus cross sections implies some simplifications of the
problem. In the description of the cross section we consider the WIMP
interacting with the point-like nucleon, either proton or neutron. On
the other hand, the nuclear charge density is mainly sensitive to the
proton distributions folded with the electromagnetic proton form
factor.

In this article we propose a methodology to get over these
simplifications. Our idea is to consider proton and neutron
distributions generated by Hartree-Fock (HF) calculations.  A work
based on these ideas, and carried out within a relativistic-Hartree
framework, has been proposed in Ref. \cite{che11}.

The effective nucleon-nucleon interactions used in 
HF calculations have been tuned to reproduce various
ground state observables, among them also the charge root mean square
(rms) radii of many isotopes. The remarkable success in describing
elastic electron scattering data, and consequently charge
distributions, indicates the reliability of these calculations in the
description of the nuclear ground states.  

In the present work we use a specific implementation of the
non-relativistic HF theory, based on a density-dependent and
finite-range nucleon-nucleon effective interaction \cite{dec80}.  The
performances of our approach have been tested against those of HF
model which uses a zero-range interaction and those of a relativistic
Hartree approach \cite{co12a}. We found that the three types of
calculations provide very similar descriptions of the ground state
properties of spherical and closed shell nuclei. For this reason we
are confident that our results are not simply related to a specific
implementation of a particular nuclear structure model, but they
represent the general predictions of any microscopic mean-field
theory.

The basic ingredients of our model are presented in
Sect. \ref{sec:model} and we show and discuss our results in Sect.
\ref{sec:res}, and in Sect. \ref{sec:con} we draw our conclusions. We
give in the Appendix the Fourier-Bessel expansion coefficients of the
proton and neutron distributions obtained in our calculations, for all
the nuclei we have considered here.

\section{The model}
\label{sec:model}
We write the probability that, in a detector with
 $N_T$ target nuclei, 
a nucleus, with A nucleons, recoils with
an energy $E_R$ after an elastic collision with a WIMP as
\beq
\frac{{\rm d}R(\er)}{{\rm d}\er} 
\, =\,  N_T\, \int_{v_{\rm min}}^\infty \, \sigma_{\chi A} (\vchi)
\, P(\er,\vchi) \, \Phi(\vchi) \, f(\vchi)
  \, {\rm d}\vchi
\,\,\,,
\label{eq:re1}
\eeq
where $\vchi$ is the velocity of the WIMP with respect to the
detector and $\vmin$ is
the minimum velocity that produces a detectable recoil in the
apparatus. In the above equation, we have indicated with $
\sigma_{\chi A}$ the WIMP-nucleus elastic cross section, with $P$ the
probability that the nucleus after been struck by the WIMP with
velocity $\vchi$ recoils with energy $\er$, with $\Phi$ the flux of
WIMPs with velocity $\vchi$, and with $f$ the probability that the
WIMP has velocity $\vchi$.

We express the number of target nuclei as the product of the detector
mass times the Avogadro's number divided by the target molecular weight, $M_d\,
N_A/A_T$.  The WIMPs flux is given by the WIMPs density times the
velocity.  We express the WIMPs density as
$\delta_\chi/\mchi c^2$, where $\delta_\chi$ is the energy density of
the WIMPs, $\mchi$ is the WIMP mass and $c$ indicates, as usual, the speed of the light.

After the collision with a WIMP of velocity $\vchi$, the value of the
nuclear recoil energy depends on the scattering angle. The
nucleus may recoil with energies values between zero and
\beq
E_{R,{\rm max}}\,= \, \frac{2 \,\muchia^2}{\ma}\, \vchi^2 
\,\,\,,
\label{eq:re2}
\eeq
where $\ma$ is the mass of the target nucleus and $\muchia = \mchi \ma
/ (\mchi + \ma)$ is the WIMP-target nucleus reduced mass.  The
energetics of the collision is such that the scattering is dominated
by the $s$ wave, therefore, because of the spherical symmetry of the
scattered wave function, the probability $P$ is independent on the scattering
angle, and can be expressed as
\beq
P (\er,\vchi) \,=\, \frac{\ma}{2\,\muchia^2} \, \frac{1}{\vchi^2}
\,\,\,,
\label{eq:P}
\eeq 
with $0 \le \er \le E_{R,{\rm max}}$.

For the evaluation of $f$ we consider a Maxwell's distribution of the
velocities centered at an average velocity $\vz$. In this way, we
obtain for the differential rate of recoil the expression
\beq
\frac{{\rm d}R(\er)}{{\rm d}\er} 
\, =\,  \frac{\delta_\chi}{\mchi c^2} \, \frac {M_d \,N_A}{A_T} \, 
\sigma_{\chi A} \, \frac{\ma \,c^2}{2 \,\muchia^2 \,c^4} \,  
\frac{2 \,c^2 \,\exp[-(\vmin / \vz)^2]} {\vz \, \sqrt{\pi} }
\,\,\,.
\label{eq:difrate}
\eeq
The total event rate, per unit of time, in a detector with energy
detection threshold $\etr$ is 
\beq
R(\etr) \, = \, 
\int_{\etr}^\infty \frac{{\rm d}R(\er)}{{\rm d}\er}  \, {\rm d}\er 
\,\,\,.
\label{eq:totrate}
\eeq

The characteristics of the detector define $M_d$ and the mass of the
recoiling nucleus. The unknown inputs of the calculations are the WIMP
mass $\mchi$, the energy density $\delta_\chi$, the average velocity
$\vz$, and the WIMP-nucleus cross section $\sigma_{\chi A}$.

The WIMP-nucleus cross section can be separated in spin-independent
(SI) and spin-dependent (SD) terms \cite{eng92}, the latter one active
only in odd-even nuclei. In this work we are only interested in the SI
term which we write as  
\beq
\sigma^{\rm SI}_{\chi A} \, =\,  \sigma_{\chi p}\, 
\left| Z \, F_p(q) \,+\, \gnp  \, N \, F_n(q) \right|^2
\,\,\,,
\label{eq:sigma2}
\eeq
where $Z$ and $N=A_T-Z$ are, respectively, the number of protons and
neutrons inside the target nucleus. In the above expression we have
factorized the elastic WIMP-proton cross section $\sigma_{\chi p}$,
and we have indicated with $\gnp=\sigma_{\chi n}/\sigma_{\chi p}$ the
ratio between the WIMP-neutron and proton cross sections.  All the
nuclear structure information is contained in the proton and neutron
form factors $F_p(q)$ and $F_n(q)$ which depend on the modulus
$q=|\bq|=\sqrt{2 \mchi \er}$ of the momentum transferred by the WIMP
to the nucleus.  The form factors are defined as the Fourier
transforms of the proton and neutron density distributions
\beq
F_{\alpha}(q) \,= \, \int \exp(i \bq \cdot \br) \, \rho_{\alpha}(r) \, {\rm d}^3 r\, = \,
 4\, \pi\, \int_0^\infty j_0(qr)\, \rho_{\alpha}(r) \, r^2 \, {\rm d}r
\,\,\,,
\label{eq:ff}
\eeq
where $\alpha=p$ for protons and $n$ for neutrons and $j_0$ indicates the zeroth-order spherical Bessel function. 

In the above equation we have already considered that in 
our calculations the densities distributions have spherical
symmetry. The expression (\ref{eq:sigma2}) of the cross section, where
the proton and neutron numbers are explicitly written, implies that the
densities should be normalized as 
\beq
\int \rho_{p,n}(r) \, {\rm d}^3 r \,= \,
 4 \,\pi\, \int_0^\infty \rho_{p,n}(r) \, r^2 \, {\rm d}r \,= \,1 
\,\,\,.
\label{eq:rho}
\eeq

In addition to these distributions we have considered the matter
distribution 
\beq \rho_m \, =  \half \,\left(  
\rho_p\,+\, \rho_n  \right)
\,\,\,,
\eeq 
and the charge distribution, $\rho_{ch}$.  The latter one is obtained
by folding the proton distribution with the electromagnetic nucleon
form factor.  The densities $\rho_m$ and $\rho_{ch}$ have been used to
calculate the form factors $F_m(q)$ and $F_{ch}(q)$ with expressions
analogous to the (\ref{eq:ff}).

As we have already pointed out in the introduction, the procedure
commonly adopted in the literature considers the form factors obtained
by using the experimental charge density distributions. This implies
the assumption that proton and neutron density
distributions coincide with the charge distributions, therefore, in
this case, instead of Eq. (\ref{eq:sigma2}) the expression
\beq
\sigma^{\rm SI}_{\chi A} \, =\,  \sigma_{\chi p}\, F^2_{ch}(q) \,
\left| Z \,+\, \gnp  \, N \right|^2
\,\,\, ,
\label{eq:sigma2red}
\eeq
is used. 

In the present work, the proton and neutron distributions have been
calculated by using a mean-field approach where the densities are
given by the expression
\beq
\rho(r)\, =\,  \sum_{k \le k_F}\, |\phi_k(\br)|^2
\,\,\,.
\label{eq:dens1}
\eeq
In the above expression $\phi_k$ indicates the single particle (s.p.)
wave function characterized by the quantum numbers identified by $k$
and the sum is limited to all the states below the Fermi surface.
Clearly, proton and neutron densities are obtained by selecting in the
sum only those s.p. states related to protons or neutrons
respectively.

In our calculations the s.p. wave functions are obtained by solving
the HF equations \cite{rin80}
\beq
-\frac {\hbar^2} {2 m} \,\nabla^2  \phi_k(\br) \,
+ \, U(\br) \, \phi_k(\br) \, -\, \int {\rm d}^3r' \, W(\br,\br') \, \phi_k(\br') 
\, = \, \epsilon_k  \, \phi_k(\br) 
\,\,\,,
\label{eq:HF}
\eeq
where $m$ is the nucleon mass, $\epsilon_k$ the s.p. energy and 
we have indicated the Hartree term as
\beq
 U(\br) \, =\,  \int {\rm d}^3r' \, V(\br,\br') \, \rho(\br')
\,\,\,,
\eeq
and the Fock - Dirac term as
\beq
W(\br,\br') \,= \, \sum_{k' \le k_F} \,V(\br,\br') \, 
\phi^*_{k'}(\br') \, \phi_{k'}(\br) 
\,\,\,.
\eeq
In the above equations, $V(\br,\br')$ represents the effective
nucleon-nucleon interaction. In our calculations we use an interaction
which has finite-range character in the scalar, isospin, spin and
spin-isospin terms and zero-range character in the spin-orbit and
density-dependent terms. This interaction, known in the literature as
Gogny interaction \cite{dec80}, contains 13 parameters whose values
have been chosen to reproduce experimental masses and charge radii of
a large number of nuclei.  We have done our calculations with two
different choices of the parameter values, the more traditional D1S
force \cite{ber91}, and the more modern D1M force \cite{gor09}.  Since
we did not find relevant differences in the results obtained with the
two effective interactions, we show here only the results obtained
with the D1M force.

In our calculations we treat spherical nuclei, therefore we found
  convenient to write the s.p. wave functions by
factorizing a term $R_{nlj}(r)$ depending on the distance $r$ from the
nuclear center and another part related to the spherical harmonics. In
this case, the density (\ref{eq:dens1}) assumes the expression
\beq
\rho(r) \, =\, \frac{1}{4 \pi} \,\sum_{nlj} \, v_{nlj}^2 \, (2j+1) \, \left| R_{nlj}(r) \right|^2
\,\,\,.
\label{eq:dens2}
\eeq
where $n,l$ and $j$ indicate, respectively, the principal quantum
number, the orbital and total angular momentum characterizing the
s.p. state whose degeneracy is $2j+1$. In Eq. (\ref{eq:dens2}) we have
indicated with $v_{nlj}^2$ the occupation probability of the
s.p. state. In closed shell nuclei we have $v_{nlj}^2=1$ for the
states below the Fermi surface, and $v_{nlj}^2=0$ for those above
it. In our calculations we have studied three nuclei of this type,
\oxy, \ca\/ and \pb\/ which are representative of various regions of the
nuclear chart.

%
\begin{table}
\begin{center}
\begin{tabular}{cccccccc}
\hline\hline
 ~ &~~&\multicolumn{2}{c}{BE/nucleon [MeV]} 
      &~& \multicolumn{3}{c}{rms [fm]} \\
      \cline{3-4} \cline{6-8}
 Nucleus &~~& HF & Exp. &~~& HF & BCS & Exp. \\
\hline
 $^{16}$O   & & 7.98  & 7.98 && 2.76 &  -   & 2.74   \\ 
 $^{40}$Ar  & & 9.32  & 8.60 && 3.40 & 3.40 & 3.42   \\ 
 $^{40}$Ca  & & 8.51  & 8.55 && 3.47 &  -   & 3.48   \\ 
 $^{72}$Ge  & & 8.57  & 8.73 && 4.02 & 4.02 & 4.05 \\
 $^{136}$Xe & & 8.44  & 8.40 && 4.80 & 4.81 &  -  \\
 $^{208}$Pb & & 7.83  & 7.87 && 5.48 &  -   & 5.50   \\ 
\hline\hline
\end{tabular}
\caption{\small 
  Binding energies (BE) per nucleon and charge rms radii
  calculated with the D1M interaction by using the HF and BCS models 
  and compared with their experimental values. 
}
\label{tab:be}
\end{center} 
\end{table}  

The performances of our calculations in reproducing ground state
observables are rather good as it is shown in Table \ref{tab:be} where
we compare binding energies and charge rms radii with their
experimental values. The good description of the ground state
properties within the HF theory is not related to the present
implementation of the mean-field approach, but it is a feature of this
type of calculations, as it is shown in Ref. \cite{co12a}, where our
results are compared with those obtained by using a zero-range
interaction and also with those produced within a relativistic
framework.

Because of their use
in WIMPS detectors \cite{aco08,aal11,ahm11}, we 
have also considered three open shell spherical nuclei, \ar, \ger\/ 
and \xe.  In the HF picture, in
these nuclei the s.p. levels are not fully
occupied. In this case, the pairing phenomena, irrelevant in
closed-shell nuclei, cannot be neglected. We have considered
pairing effects by solving the equations of the Bardeen, Cooper and
Schriefer (BCS) theory applied to finite nuclear systems
\cite{rin80}. 
We start from
the s.p. wave functions and energies generated by our HF calculations,
and then we solve the set of equations:
\beq
\Delta_a\, = \, -\frac{1}{2 \sqrt{2j_a+1}} 
\, \sum_b \, \sqrt{2j_b+1} \, u_b \, v_b \, \langle bb0 | V | aa0 \rangle
\,\,\,,
\label{eq:gap1}
\eeq
\beq
N \, = \,  \sum_a \, (2 j_a+1) \, v_a^2 \, = 
\,  \sum_a \, \left(j_a+\half \right) \,
\left\{         1 - \frac{\epsilon_a - \lambda} 
               { \sqrt{(\epsilon_a - \lambda)^2 + \Delta^2_a} }
\right\}   
\,\,\,.
\label{eq:gap2}
\eeq
In the above equations we have used $u^2=1-v^2$, 
$a$ and $b$ indicate different sets of the $n,l,j$ quantum
numbers, and N is the total number of nucleons subject to the
  action of the pairing.
The matrix element of the interaction is calculated by coupling to
zero the angular momenta of the $a$ and $b$ s.p. states.  The solution
of these equations provides the values of $\lambda$ which is the gap
energy, i. e. the energy difference between the last occupied level
and the first empty level in the HF picture.  The other output of the
calculation, i. e. the values of the $v_a$, is more relevant for our
purpose since, as expressed by Eq.  (\ref{eq:dens2}), it modifies the
density distributions.

We carried out the BCS calculations by using the same nucleon-nucleon
interaction used in the HF calculations, but without Coulomb and
spin-orbit terms, as it is commonly done in the literature for this
type of calculations. 
The relevance of the pairing
can be deduced by observing how much the values of $v^2$ differ from
those of the HF picture. We show in Table \ref{tab:v2} the values of
$v^2$ related to the partially occupied s.p. states nearby the Fermi
level, for the three open shell nuclei we have considered. The values
of $v^2$ for the neutrons in the \xe~ nucleus clearly indicate the
shell closure for N=82.  

%
\begin{table}[h]
\begin{center}
\begin{tabular}{ccccccc}
\hline \hline
&&  \multicolumn{5}{c}{$v^2$} \\
\cline{3-7} 
 Nucleus &~~& \multicolumn{2}{c}{protons} 
       &~~& \multicolumn{2}{c}{neutrons} \\
\hline
 $^{40}$Ar && $1d_{3/2}$ & $1f_{7/2}$ && $2p_{3/2}$ & $1f_{7/2}$ \\
     && 0.655  & 0.008  && 0.003 & 0.257 \\
\hline
 $^{72}$Ge && $2p_{3/2}$ & $1f_{5/2}$ && $2p_{1/2}$ & $1g_{9/2}$ \\
     && 0.541 & 0.286 && 0.984 & 0.013 \\
\hline
 $^{136}$Xe && $1g_{7/2}$ & $2d_{5/2}$ && $1h_{11/2}$ & $1h_{9/2}$ \\
      && 0.388 & 0.115 & &1.000 & 0.000 \\
\hline \hline
\end{tabular}
\caption{\small 
Values of $v^2$ for the partially occupied s.p. levels near the Fermi
level.  
}
\label{tab:v2}
\end{center} 
\end{table}

\section{Results}
\label{sec:res}

We have calculated proton, neutron and charge density distributions
within the theoretical framework outlined in the previous section.
The charge distributions have been obtained from the proton
distributions by folding them with the proton electric form factor
extracted from elastic electron-proton data. The results we show here
have been obtained by using a dipole parametrization of the proton
form factor \cite{bof96}. We have verified that more accurate
parametrizations \cite{hoe76} modify our charge distributions on few
parts on a thousand. 

\begin{figure}[b]
\begin{center}
\includegraphics[scale=0.4]{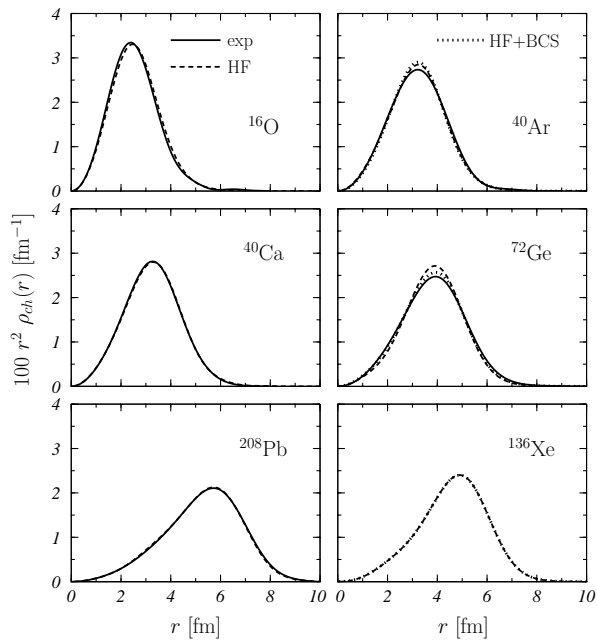} 
\caption{\small Charge distributions obtained with our HF (dashed
  lines) and HF+BCS (dotted lines) calculations compared with the
  empirical distributions (full lines) \cite{dej87}, extracted from
  elastic electron scattering data. The distributions, normalized as
  indicated by Eq. (\ref{eq:rho}), have been multiplied by $r^2$.  }
\label{fig:charge}
\end{center}
\end{figure}
\begin{figure}
\begin{center}
\includegraphics[scale=0.4]{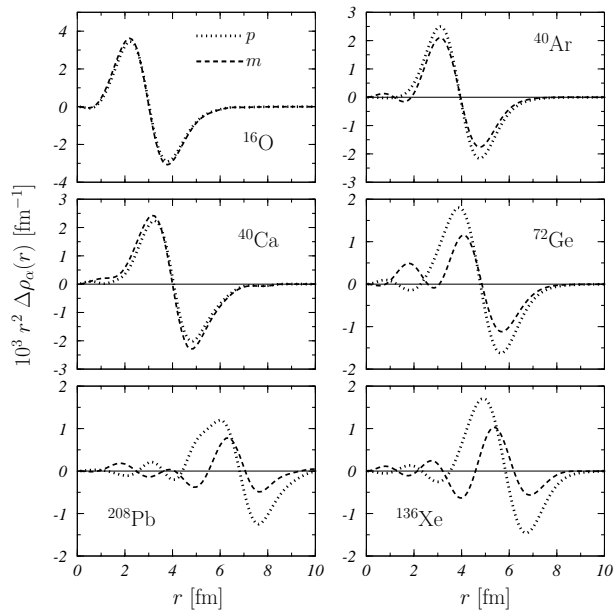} 
\caption{\small 
Differences between proton and charge distributions (dotted lines) and
between matter and charge distributions (dashed lines), 
Eq. (\ref{eq:drho}), multiplied by $r^2$.
  }
\label{fig:ddens}
\end{center}
\end{figure}

In Fig. \ref{fig:charge} we compare the charge distributions obtained
in our HF and HF+BCS calculations with the empirical ones
\cite{dej87}, extracted from elastic electron scattering experiments.
In this figure we present the distributions multiplied by $r^2$ to
directly show the functions which are integrated in Eq. (\ref{eq:ff}).
We observe a very good agreement of our results with the experimental
distributions.  This is a consequence of the fact that the charge rms
radii are some of the experimental data included in the fit procedure
used to define the parameters of the D1M force. As it is well known,
the main differences between experimental and theoretical charge
distributions arise in the nuclear interior. In general, the
experimental charge distributions are smoother than the theoretical
ones. A well known example of these phenomena is the case of \pb\/
which has been widely investigated from both 
experimental \cite{cav82} and theoretical \cite{co87c,ang01} points of
view.

For the open shell nuclei, we compare the results obtained in both HF
and HF+BCS frameworks, dashed and dotted lines, respectively.  For the
two nuclei where the empirical charge distributions are available,
nuclei the HF+BCS distributions provide a better description of the
experimental ones than those obtained in the HF calculation.  This is
more evident in the case of the \ger\/ nucleus.

The results of Fig. \ref{fig:charge} provide an indication of the
capacity of our model to describe the charge distributions. The
WIMP-nucleus interaction is active on the full matter distribution
which we claim to describe with the
same accuracy than the charge distribution.  Since no accurate data on
matter distributions are available we 
cannot test this assumption, which, on the other hand, we
believe it is well grounded.

In the literature, properly normalized charge distributions are used to
estimate WIMP-nucleus cross sections instead than matter
distribution. In order to estimate the error done in this
approximation we have calculated the difference between proton ($\alpha=p$) and
matter ($\alpha=m$) distributions with respect to the charge distributions
\beq
\Delta \rho_{\alpha}(r) \,=\,  \rho_{\alpha} (r) \,- \, \rho_{ch} (r) 
\,\,\, .
\label{eq:drho}
\eeq
We show in Fig. \ref{fig:ddens} these
differences, multiplied by $r^2$, obtained in the HF framework for the
\oxy, \ca\/ and \pb\/ nuclei, and in the HF+BCS framework for the \ar,
\ger\/ and \xe\/ nuclei. Unless stated otherwise, the results presented
henceforth have been obtained in this manner.

The first remark about the results of Fig. \ref{fig:ddens} is that
$\Delta \rho_{p}(r) $ and $\Delta \rho_{m}(r)$ have the same order of
magnitude in all the nuclei considered.  This indicates that the
contribution of the electromagnetic proton form factor is not
negligible. In the opposite case, we would have found smaller
differences with the proton distributions than with the matter
distributions.
A detailed discussion on the WIMP-nucleon interaction is
not in the scope of the present article. We point out, however, that in the
expression (\ref{eq:sigma2}) of the cross section, the WIMP-proton
cross section is factorized.  
From the theoretical point of view,
this implies the requirement of using point-like densities to calculate
the form factors $F_{p}(q)$ and $F_{n}(q)$ in order to avoid a kind of double
counting between electromagnetic and weak nucleon form factors.

A second remark concerning the results shown in Fig. \ref{fig:ddens}
indicates that in nuclei with equal number of protons and neutrons,
\oxy\/ and \ca, the proton and matter distributions, when normalized as
indicated in Eq. (\ref{eq:rho}), are very similar. The differences
start to appear in the \ar\/ nucleus where $Z=18$ and $N=22$. The larger
is the difference between $Z$ and $N$, the larger is the difference
between proton and matter distributions. In \ger, \xe\/ and \pb\/ nuclei
$\Delta \rho_p$ and $\Delta \rho_m$ differ mainly in the nuclear
interior.  In this region, the differences with the charge
distributions, here considered as reference result, are even out of
phase. The situation becomes more stable in the surface region where
the curves show similar behaviors. The negative values of all the
curves in the surface region is due to the fact that the charge
distributions are always wider in space than the point-like
distributions. This is the effect of the folding with the proton
electromagnetic form factor. Interesting to notice that in all the
cases, with the exceptions of \oxy\/ and \ca, the $\Delta \rho_p$ is
always larger than $\Delta \rho_m$. Our calculations indicates that
the identification of the charge distributions with the full matter
distributions is a better approximation than that obtained by
identifying charge with proton distributions.

\begin{figure}
\begin{center}
\includegraphics[scale=0.4]{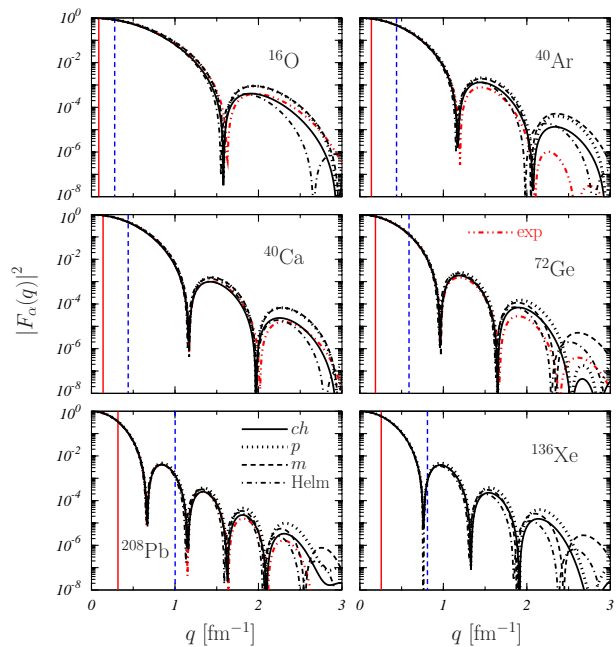} 
\caption{\small (Color on line) 
Nuclear form factors calculated with various density
distributions. Full, dotted and dashed lines show the results obtained by using,
respectively,
the charge, proton and matter distributions obtained in our
calculations.
The red dashed-doubly-dotted 
lines are the form factors obtained by
using the empirical charge distributions. The dashed-dotted  lines indicate 
the form factors obtained within the Helm model. 
The vertical blue lines indicate the value of the momentum transfer
corresponding to a nucleus kinetic energy of 100 keV. 
}
\label{fig:ffq}
\end{center}
\end{figure}
\newpage
\begin{figure}
\begin{center}
\includegraphics[scale=0.4]{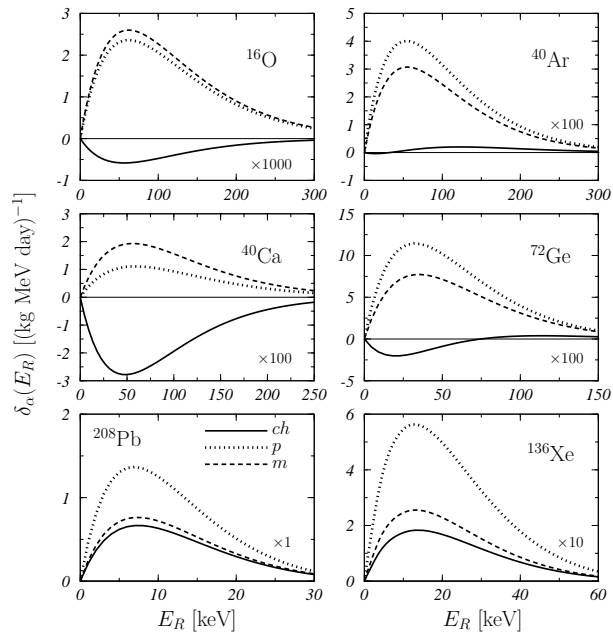} 
\caption{\small 
Differences $\delta_\alpha(\er)$, as defined in Eq. (\ref{eq:dder}), between the differential rates (\ref{eq:re1})
calculated with our density distributions and those obtained 
with the Helm form factors. 
The values for the charge (solid lines), proton
(dotted lines) and matter (dashed lines) distributions are shown. 
In each panel, the multiplication factors used to enlarge the y
  scale are indicated.
}
\label{fig:drate}
\end{center}
\end{figure}

The form factors to be used in Eq. (\ref{eq:sigma2}) to calculate the
WIMP-nucleus cross section are shown in Fig. \ref{fig:ffq} where we
have indicated, respectively, 
with full, dotted and dashed lines the results obtained by using
the charge, proton and matter distributions obtained in our
calculations. In the figure we also
show with the red dashed-
doubly-dotted lines the form factors obtained by
using the empirical charge distributions. The dashed-dotted 
lines show the Helm form factors \cite{hel56} which are commonly used
in the literature to calculate the WIMP-nucleus cross section.  The
Helm form factor is obtained by considering a charge density with a
Gaussian surface distribution, and has a simple analytic expression
\beq
|\fhelm(q)|^2 \, =\,  \left[ \frac{3\, j_1(q R)}{q R}\right]^2 \, 
\exp(- q^2 s^2)
\,\,,
\label{eq:helm}
\eeq 
where $j_1$ is the first-order spherical Bessel function, and $R$ and $s$
are two parameters which are chosen to reproduce at best the
experimental charge density distributions on all the nuclear chart. We
used the parametrization suggested in Refs. \cite{lew96,che11} 
\beqn
R^2 &=& c^2  \,-\, 5 s^2 \,+\, 18.65 \,{\rm fm}^2 \,\,\,,\\ 
c &=& \left( 1.23 \, A^{1/3} \,-\, 0.6\, \right) {\rm fm} \,\,\,, \\ 
s &=& 0.9 \, {\rm fm} \,\,\, .  
\eeqn

The differences between the various form factors increase with the value of
$q$. This reflects the fact that, by increasing the resolution power,
the probe is able to perceive the differences between the various
densities. The form factors drop rather quickly with increasing value
of $q$. The relevance of these differences in the calculation of the
total event rate can be estimated by considering that, in the
expression (\ref{eq:totrate}), the integration starts from the value
of the threshold detection energy $\etr$. As a reference example, in
Fig. \ref{fig:ffq} we have indicated with vertical blue lines the
values of the momentum transfer corresponding to 
a recoil energy, $\er$, of 100
keV. This is a large value with respect to the performances of modern
detectors. In this example, the form factors to the left hand side of
the vertical lines do would
not contribute to the total rate. This indicates
that the largest values of the form factors are excluded, confirming
an obvious consideration, the lower is the detection energy threshold,
the higher is the rate of the detected events. More interesting is the
  comparison between the various nuclei. The heavier is the nucleus, the
  higher is the value of $q$ required to obtain a 100 keV recoil
  energy. For the two heavier nuclei \xe\/ and \pb, the lines appear after
  the first diffraction minimum, therefore, for these nuclei, the
  differences between the various densities can be more relevant.

  We used our density distributions to calculate the differential, and
  total event rates, Eqs. (\ref{eq:re1}) and (\ref{eq:re2}). The values
  of the input parameters have been chosen coherently with what is used
  in the literature. If not stated otherwise, the results of the
  calculations we present in this paper have been obtained by assuming
  $\sigma_{\chi p} =$
  10$^{-42}$~cm$^2$ = 10$^{ -16}$ fm$^2$,
  $\delta_\chi=0.3$ GeV/fm$^3$, $v_0=0.001\, c$,  $g_{np}=1$, $M_\chi = 100$
  GeV. 

  The first results we want to discuss are related to the differential
  rates of eq. (\ref{eq:re1}). We show in Fig. \ref{fig:drate} the differences
\beq
\delta_\alpha(\er) \, = \, \displaystyle \left[ \frac{{\rm d}R(\er)}{{\rm d}\er} \right]_\alpha \,- \, \left[\frac{{\rm d}R(\er)}{{\rm d}\er}\right]_{\rm Helm} 
\,\,\,,
\label{eq:dder}
\eeq
  between the differential rates obtained with the distributions found
  in our calculations and those obtained by using the Helm form
  factors as a function of the recoil energy $\er$. In these
  calculations we used a threshold detection energy of 10 eV only.
  Full, dotted and dashed lines show the differences obtained by using
  the charge ($\alpha=ch$), proton ($\alpha=p$) and matter
  ($\alpha=m$) densities, respectively.

  A first observation is that these differences are rather small.  The
  relative differences in the peaks of the figure are 5\% at most.  A
  second observation is that these differences increase with
  increasing mass number. The order of magnitude of these differences
  in \pb\/ is 1000 times larger than that of \oxy, 100 times larger
  than those of \ca, \ar\/ and \ger\/ and, only, 10 times larger than
  that of \xe.  In all the nuclei we have considered, with the
  exception of \ca, the smallest differences we found are those with
  the charge distributions. This could be expected since the
  parameters of the Helm form factors have been chosen to describe
  charge distributions. For the nuclei with $Z \ne N$ the results
  obtained with the proton distributions differ more than those
  obtained with the matter distributions. This is a direct consequence
  of what we have observed in Fig. \ref{fig:ddens}.

\newpage
\begin{figure}
\begin{center}
\includegraphics[scale=0.4]{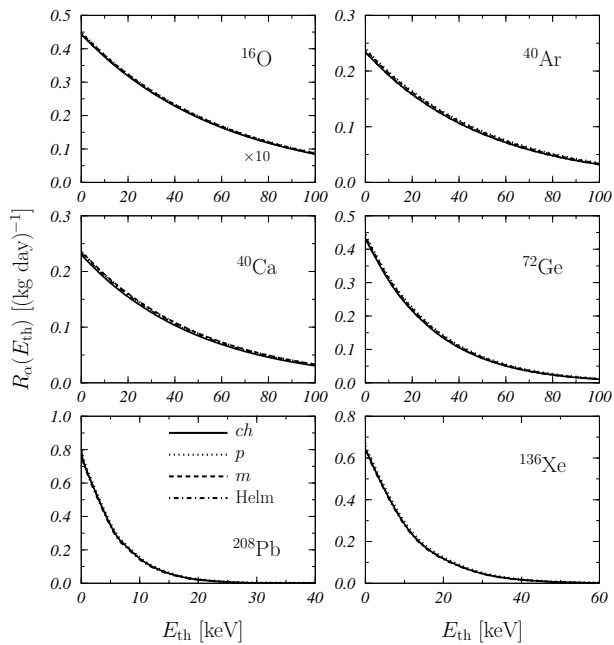} 
\caption{\small Total rates, as given by Eq. (\ref{eq:totrate}), as a
  function of the threshold detection energy. The values found for the
  charge (solid lines), proton (dotted lines) and matter (dashed
  lines) distributions as well as those obtained 
  with the Helm form factors (dashed-dotted
  lines) are shown. The values of the \oxy\/
  nucleus have been multiplied by 10.  }
\label{fig:trate}
\end{center}
\end{figure}
\newpage
\begin{figure}
\begin{center}
\includegraphics[scale=0.4]{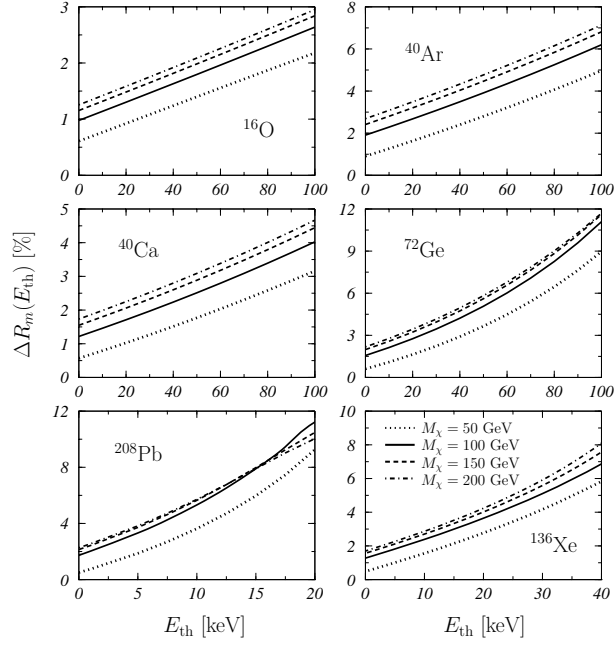} 
\caption{\small 
Relative differences, as defined by Eq. (\ref{eq:delta}), between total rates calculated with matter distributions within our model
and those obtained with the Helm form factors. 
The different lines show the results obtained by using different 
values of $\mchi$. 
}
\label{fig:diff}
\end{center}
\end{figure}
\newpage
\begin{figure}
\begin{center}
\includegraphics[scale=0.4]{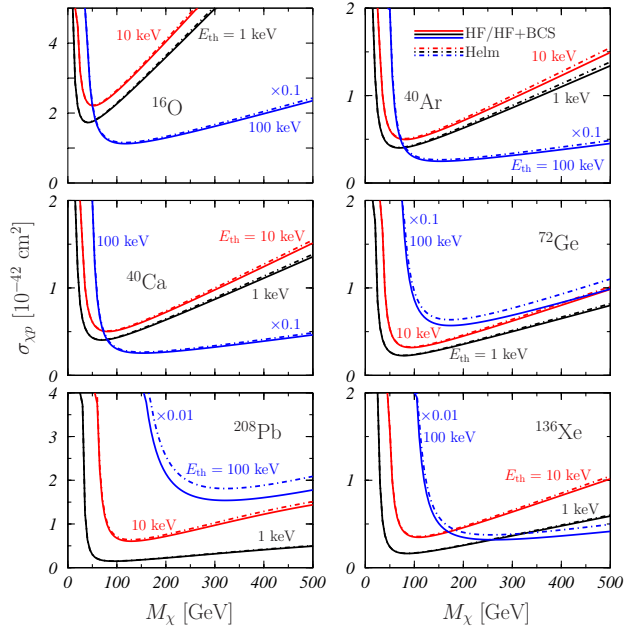} 
\caption{\small Upper limits of the WIMP-proton
  cross section, $\sigma_{\chi p}$, as a function of $M_\chi$, for
  the detection of a
  single event per day identified in
  an ideal detector of 100 tons, for $\etr=1$, 10 and 100
  keV. The solid lines show the results obtained by using our matter
  distributions obtained in HF or HF+BCS calculations, while the
  dashed-dotted lines  indicate
   the results obtained by using the Helm
  form factors. The $\etr=100$ keV results have been multiplied by 0.1
  in case of $^{16}$O, $^{40}$Ca, $^{40}$Ar, $^{72}$Ge nuclei, and by
  0.01 for $^{136}$Xe and $^{208}$Pb nuclei.  }
\label{fig:uplim}
\end{center}
\end{figure}

  We show in Fig. \ref{fig:trate} the total rates 
 $R_\alpha(E_{\rm th})$ given by Eq. (\ref{eq:totrate}) as a
  function of the threshold energy. The different lines indicate the
  results obtained with the various densities. The behavior of the
  various lines is well understood. By increasing the value of the
  threshold energy, which means a lowering of
the sensitivity of the
  detector, the number of the events becomes smaller. The comparison of
  the results obtained for the various nuclei indicates that at low
  values of $\etr$ the heavier nuclei are more efficients in detecting
  events, because of the larger number of target nucleons inside the
  nucleus. We point out that, in the figure, the \oxy\/ results have been
  multiplied by a factor 10. On the other hand, the total rates drop
  more quickly in heavier nuclei than in lighter ones. At 40 keV the
  total rates expected in \pb\/ are much smaller than those expected in
  the medium-heavy nuclei we have considered, \ca, \ar\/ and \ger.

  The scale of Fig. \ref{fig:trate} does not allow to appreciate the
  differences between the various results. For this reason, we have
  calculated the relative differences between the total rates obtained
  by using the matter distributions with respect to the results obtained
  with the Helm form factors 
  \beq
  \Delta R_m(\etr)\, = \,
   \frac { R_m(\etr) \,- \, R_{\rm Helm}(\etr)}
         {R_{\rm Helm}(\etr)}
  \,\,\,.
  \label{eq:delta} 
  \eeq
  These relative differences are shown in Fig. \ref{fig:diff} as a
  function of $\etr$. In this figure, in addition to the results
  related to total rates presented in Fig. \ref{fig:trate}, we show
  also results obtained by changing the WIMP mass $M_\chi$. 

  The value of the relative differences increases the heavier is the
  nucleus. We obtain a maximum value of 3\% in \oxy\/ and 10-20\% in
  \pb. The general enhancement with increasing $\etr$ is more related to the
  lowering of the value in the denominator of Eq. (\ref{eq:delta}) than
  to a real increase of the difference between the results obtained with 
  matter and Helm distributions.  

  By comparing the results obtained with different values of $M_\chi$,
  we observe that $\Delta R$ increases with increasing mass.
  Since the minimum value of the momentum transfer for the detection is
  $q_{\rm min} = \sqrt{2\, \etr \, M_\chi}$, if $M_\chi$ increases, also
  $q_{\rm min}$ increases. Therefore, a larger part of the form factors
  at low $q$, shown in Fig. \ref{fig:ffq}, is excluded from the integral
  of Eq. (\ref{eq:re2}), and the final result is more sensitive to the
  differences between various calculations, differences which appear at
  high $q$ values.  In heavier nuclei, the increase of $\etr$ moves
  the limit of this integration nearby the first minimum of the form
  factor, which is slightly displaced in the Helm model with respect to
  that obtained with the matter distribution. 

  To estimate the relevance of the differences in the use of the
  various form factors, we have calculated hypothetical values of the
  WIMP-proton cross section for which a rate of a single detected
    event  per day is obtained in a 100
  ton detector.  For all the nuclei we have investigated, we show
  in Fig. \ref{fig:uplim} these limits, obtained for
    $\etr=1$, 10 and 100 keV, as a function of the WIMP mass.
The
  dashed-dotted curves correspond to the values found with the Helm
  form factors, while the solid lines 
  indicate, respectively,  those obtained by using our
  HF (for \oxy, \ca\/ and \pb) and HF+BCS (for \ar, \ger\/ and
    \xe) matter distributions 
and HF+BCS (for \ar, \ger\/ and \xe). The $\etr=100$ keV
  results have been multiplied by the factors indicated.

  Obviously, the lower is $\etr$, the lower is the line of the exclusion
  plot.  The differences between our results and those obtained with the
  Helm form factors become larger with increasing target mass, with
  increasing $M_\chi$ values, and with increasing $\etr$. These results
  show a minimum difference of 1.4\% in the case of \oxy\/ for $\etr = 1$
  keV, and a maximum one of 15\% for \pb\/ with $\etr = 100$ keV. 

  The results we have so far presented have been obtained by assuming
  the same coupling strength of the WIMP with protons and neutrons,
  i. e. $\gnp=1$ in Eq. (\ref{eq:sigma2}). We expect that, by
  releasing this assumption, as suggested, for example, in
  Ref. \cite{far11}, the requirement of a correct description of
  proton and neutron distributions becomes more important. For this
  reason, we have carried on calculations for different values of
  $\gnp$.  We have chosen values of $\gnp$ producing extreme
  situations. The largest and smallest values we have considered,
  $\pm2$, slightly exceed those 
  indicated in \cite{far11},
  values based on a
    compatibility analysis of the DAMA \cite{ber08,ber10}, CoGeNT
  \cite{aal11} and XENON100 \cite{apr11} data. The other values of
  $\gnp$ have been chosen because of their physical meaning. The
  isospin conserving coupling, which has been our reference up to now,
  is obtained with $\gnp=1$. The value $\gnp=0$ selects only the
  WIMP-protons interaction. Finally, by using $\gnp=-1$ we generate a
  cross section sensitive only to the differences between proton and
  neutron distributions.
  
\begin{figure}[h]
\begin{center}
\includegraphics[scale=0.4]{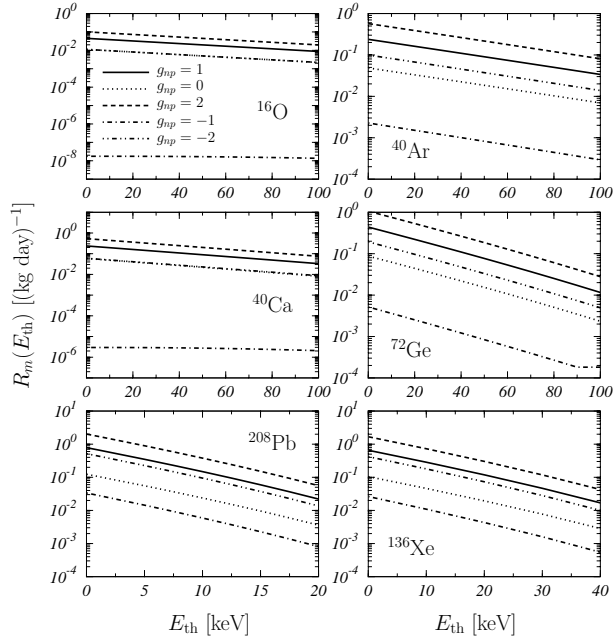} 
\caption{\small 
Total rates, Eq. (\ref{eq:totrate}), calculated for different values of
$\gnp$,  which are indicated in the insert of the \oxy\/ panel.
}
\label{fig:fnfp}
\end{center}
\end{figure}

\begin{figure}[ht]
\begin{center}
\includegraphics[scale=0.4]{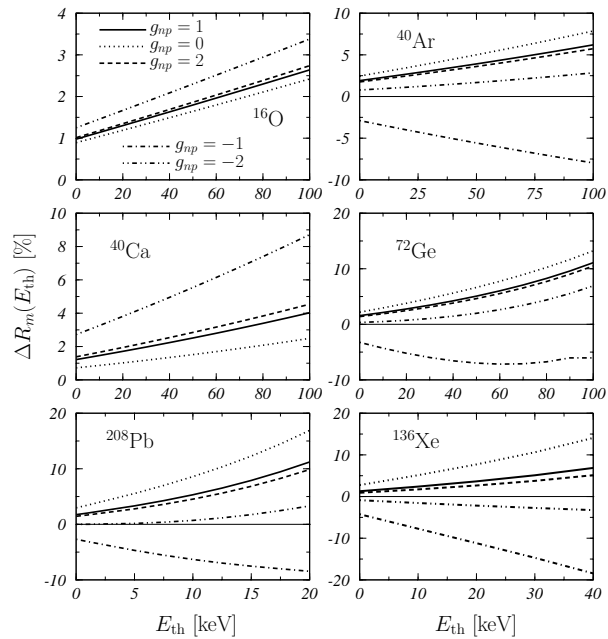} 
\caption{\small 
Relative differences between total rates calculated with our matter
distributions and those obtained with the Helm form factor
for $\mchi=100$ GeV and for different values of $\gnp$, 
indicated in the insert of the upper left panel. 
}
\label{fig:difffnfp}
\end{center}
\end{figure}
\begin{figure}
\begin{center}
\includegraphics[scale=0.4]{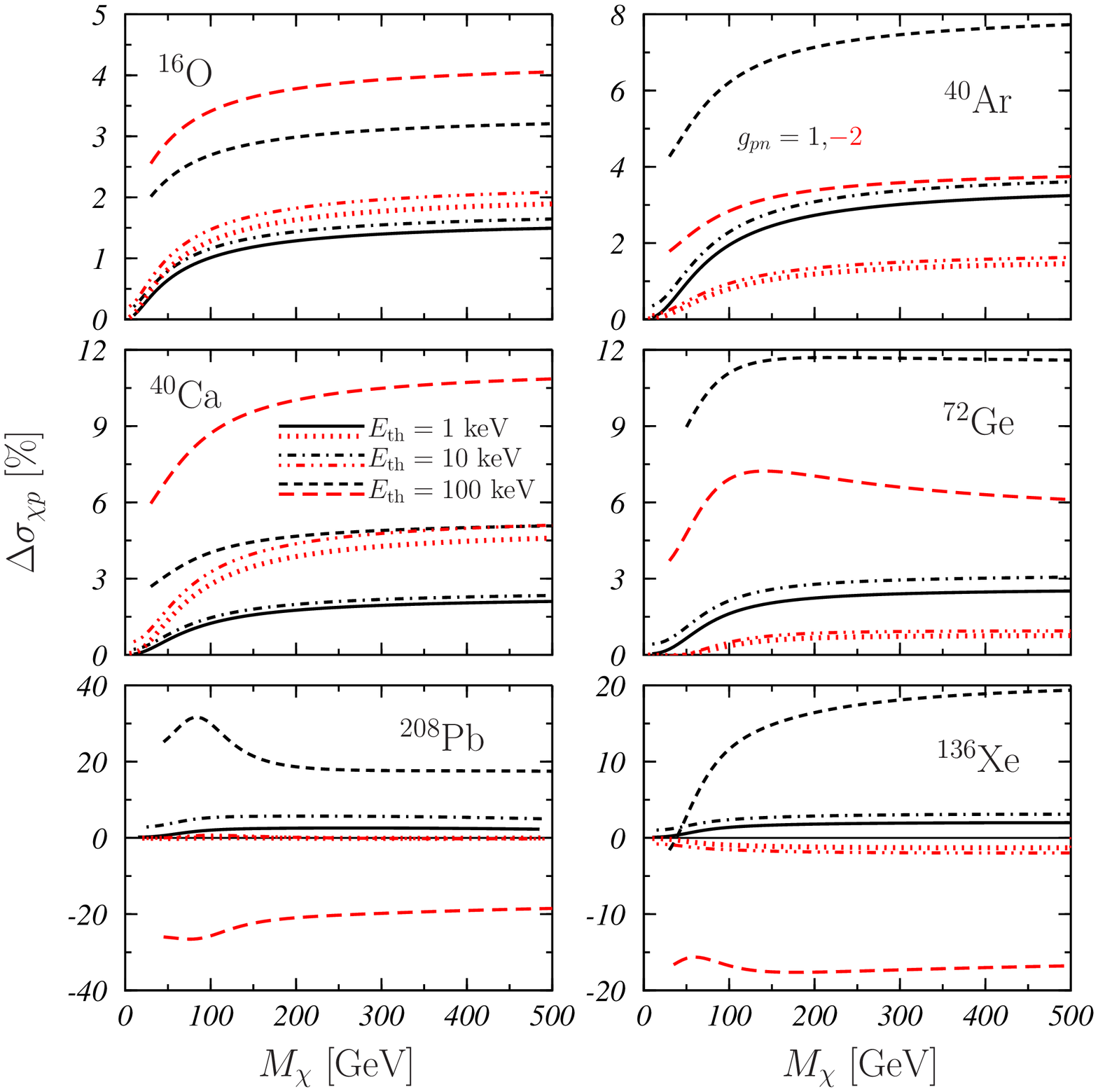} 
\caption{\small (Color on line) Relative differences between the 
  limit values of $\sigma_{\chi p}$ calculated by using our matter distributions and those
  obtained with the Helm form factors. 
  The results for $E_{\rm th}=1$ (full and red dotted lines), 10 (dashed-dotted and red dashed-double-dotted lines) and 100 keV (short-dashed and red long-dashed lines) are shown. Black full, dashed-dotted and short-dashed lines correspond to $\gnp=1.0$, while red dotted, dashed-double-dotted and long-dashed lines correspond to the isotope violating hypothesis with $\gnp=-2.0$.  }
\label{fig:diffup}
\end{center}
\end{figure}

  The total rates calculated with these values of $\gnp$, by using
  $M_\chi=100$ GeV, are shown in Fig. \ref{fig:fnfp}. In all the
  calculations the sequence of the various responses is similar.  The
  lower lines are obtained with $\gnp=-1$ the situation where the proton
  and neutron contribution cancel with each other. In the calculations
  where $F_p=F_n$, for example those which use the Helm form factors,
  the total rates for \oxy\/ and \ca\/ are exactly zero. Next we have the
  results with $\gnp=0$ where the WIMP interacts only with the proton.
  We have then the results with $\gnp=-2$ where the WIMP-neutron
  interaction dominates. In nuclei with $N=Z$ the total rates almost
  overlap with that obtained with $\gnp=0$. For positive values of
  $\gnp$ the total rates increase with increasing $\gnp$. In
  Ref. \cite{far11} a value of about -1.5 has been suggested.

  The aim of our work is to study the need of using accurate proton and
  neutron distributions to describe the WIMP-nucleus cross section. For
  this reason, we show in Fig. \ref{fig:difffnfp} the relative differences
  $\Delta R$, Eq.  (\ref{eq:delta}), obtained with the various
  values of $\gnp$.  In the two $N=Z$ nuclei, \oxy\/ and \ca, the results
  with $\gnp=-1$ are not indicated since in this case $R_{\rm Helm}$ is zero
  with the obvious consequences in the calculation of $\Delta R$.

   The behavior of the various lines has different characteristics for
  the $N=Z$ nuclei and for the other ones. In the former cases the
  smaller differences appear for $\gnp=0$ and the largest ones for
  $\gnp=-2$. The relative differences can reach values around the 20\%
  in heavy nuclei.

  We have evaluated the consequences of these differences by
  calculating limits as those shown in Fig. \ref{fig:uplim} with
  different values of $\gnp$. We show in Fig. \ref{fig:diffup} the
  relative differences $\Delta \sigma_{\chi p}$ between the limits of
  the WIMP-proton cross section calculated with our matter
  distributions and those obtained with the Helm form factors.  The
  results for $E_{\rm th}=1, 10, 100$ keV are shown.  The black (full,
  dashed-dotted and short-dashed) lines show the relative differences
  of our reference calculations done with $\gnp=1$, i.e. those
  obtained from the results of Fig. \ref{fig:uplim}. The red (dotted,
  dashed-doubly-dotted and log-dashed) lines indicate the differences
  obtained with $\gnp=-2$, a value comparable with that suggested in
  Ref. \cite{far11}.  In all the cases the largest differences have
  been obtained for a threshold energy of 100 keV. As it has been
  previously discussed, the differences are enhanced with increasing
  $\etr$. We do not identify common trends in this figure. Sometime
  the $\gnp=-2$ results show the largest differences, in other cases
  the largest differences are produced by $\gnp=1$. We observe that in
  \ar, \ger\/ and \xe\/ these values can reach the 20\%.

  \section{Conclusions}
  \label{sec:con}
  In this work, we have studied the validity of the approximation
  commonly done in the literature consisting in 
  considering the nuclear charge density instead
  than the matter density in the calculation of the WIMP-nucleus cross
  sections.  While charge distributions are experimentally known for
  various nuclei with high accuracy \cite{dej87}, the matter
  distributions are essentially unknown. For this reason, we propose
  to use the point-like matter distributions obtained in
  mean-field calculations of HF type.  We present here the results
  obtained with a specific implementation of the HF calculations,
  involving an effective finite range interaction of Gogny type
  \cite{dec80}. Our experience \cite{co12a} indicates that these
  results are general and more
  related to the mean-field description of
  the nucleus rather than to its specific implementation.

  We have shown that our model describes well the experimental charge
  distributions, and we have assumed that it is also able to properly
  describe the matter distributions.  We have compared the form
  factors calculated with our matter distributions with 
the
  Helm form factors commonly adopted in the
  literature. The differences between the factors calculated by using
  different distributions show up at high values of the momentum
  transfer, usually after the first diffraction minimum. The relevance
  of  these differences  is
  strictly related to the value of the detection threshold energy
  $\etr$, and decreases with it.  

  We have calculated differential, total rates and upper detections
  limits. In this last case, we found that the differences between our
  results and those obtained with the Helm form factor become larger
  with increasing target mass, with increasing $\mchi$ values, and
  with increasing $\etr$. We found a maximum relative difference of
  about 15\% for \pb\/ with $\etr = 100$ keV.

  The requirement of using matter instead than charge distributions in
  the calculation of the WIMP-nucleus cross section becomes more
  important when the assumption of isospin conserving WIMP-nucleon
  interaction is released. In this case, the proton and neutron
  distributions are mixed with different weights, therefore an accurate
  description of the two distributions is mandatory. Also in this case,
  the value of $\etr$ is the parameter which dominates the
  value of the uncertainty, however, the use of mixing values close to
  those suggested in the literature increases the difference with respect
  the results obtained with the Helm model. In specific situations the 
  relative differences can reach the value of 20\%. 

  In the estimation of the total event rate related to the WIMPs
  detection, the values of many input quantities are unknown, and strong
  assumptions are done on them. On the other hand, the nuclear physics
  part of the process is well understood, and we propose to use the 
  modern nuclear structure results to improve its description.

  \section*{Appendix: Fourier-Bessel coefficients of density distributions}  
  \label{sec:appa}
  In this appendix we show the values of the Fourier-Bessel coefficients
  which allow the reconstruction of our proton and neutron density
  distributions for the nuclei we have considered. These distributions
  can be obtained by the expression 
  \beq
  \rho(r) \,=\, \sum_{\mu=1}^{M} A_\mu \, \frac{\sin(q_\mu r)}{q_\mu r}
  \,\,\,,
  \label{eq:fb}
  \eeq
  where the  $q_\mu$ are defined as
  \[
  q_\mu= \mu \frac{\pi}{R} 
  \,\,\,,
  \]
  with $R$ the maximum value of the radius where the density is
  supposed to be different from zero. The density distributions obtained
  with the coefficients given in Tables \ref{tab:fb1} and \ref{tab:fb2},
  have the usual nuclear physics normalizations, i. e. the integrals 
  of Eq. (\ref{eq:rho}) are normalized to the number of protons or
  neutrons. 

%
\begin{table}
\begin{center}
{\scriptsize
\begin{tabular}{rccccccccc}
\hline \hline 
 ~ &~~ & \multicolumn{2}{c}{$^{16}$O}  
    &~~   & \multicolumn{2}{c}{$^{40}$Ca}  
     &~~  & \multicolumn{2}{c}{$^{208}$Pb} \\
\cline{3-4} \cline{6-7}  \cline{9-10}      
$\mu$     &  & proton    &      neutron    &   & proton    &      neutron    & & proton    &      neutron    \\ 
\hline
     1 &&  2.89514E-02 &  2.90605E-02 &&  6.17474E-02 &  6.23662E-02 &&  7.69622E-02 &  1.16349E-01 \\
     2 &&  5.44734E-02 &  5.53342E-02 &&  5.75764E-02 &  6.08146E-02 &&  1.87236E-02 &  2.73910E-02 \\
     3 &&  2.37463E-02 &  2.50485E-02 && -2.94985E-02 & -2.82238E-02 && -6.54988E-02 & -8.28045E-02 \\
     4 && -1.51728E-02 & -1.46814E-02 && -1.93097E-02 & -2.10479E-02 &&  3.20906E-02 &  4.81394E-02 \\
     5 && -1.94164E-02 & -1.99175E-02 &&  1.87819E-02 &  1.85152E-02 &&  1.42479E-02 &  1.17763E-02 \\
     6 && -5.40039E-03 & -6.14889E-03 &&  9.02298E-03 &  9.93418E-03 && -2.81145E-02 & -2.99372E-02 \\
     7 &&  1.85720E-03 &  1.34614E-03 && -1.89926E-03 & -2.77723E-03 &&  1.43845E-02 &  1.88786E-02 \\
     8 &&  6.24424E-04 &  3.75985E-04 &&  1.40503E-03 & -1.48915E-03 &&  1.53527E-02 & -6.90422E-03 \\
     9 && -1.18242E-03 & -1.29026E-03 &&  2.91761E-03 & -1.02014E-03 &&  3.00274E-03 & -2.34458E-02 \\
    10 && -9.50609E-04 & -9.80391E-04 &&  1.34965E-03 & -2.77186E-03 &&  3.76239E-03 & -1.08603E-03 \\
    11 && -1.71403E-04 & -1.66395E-04 &&  1.77650E-03 & -2.38989E-03 && -4.27147E-03 &  5.77452E-03 \\
    12 &&  4.61957E-05 &  6.54090E-05 &&  1.68542E-03 & -1.69061E-03 && -3.33174E-03 &  2.14901E-03 \\
    13 && -1.93572E-05 & -9.66199E-06 &&  1.13150E-03 & -8.69682E-04 &&  3.06557E-03 & -8.57191E-04 \\
    14 && -7.29176E-05 & -6.72835E-05 &&  2.12233E-04 & -4.41138E-04 &&  1.80972E-03 & -3.12159E-03 \\
    15 && - &  - &&  2.61156E-04 &  4.04638E-05 && -3.10532E-04 &  4.46753E-04 \\
\hline
R [fm] && 7.0  &  7.0 && 7.0 &  7.0 && 10.0 & 10.0 \\
\hline \hline 
\end{tabular}
}
\caption{\small Fourier-Bessel coefficients $A_\mu$ and radii $R$, see
  Eq. (\ref{eq:fb}), for the proton and neutron densities obtained in
  the HF calculations. 
  }
\label{tab:fb1}
\end{center} 
\end{table}

\begin{table}
\begin{center}
{\scriptsize
\begin{tabular}{rccccccccc}
\hline\hline
 ~ &~~ & \multicolumn{2}{c}{$^{40}$Ar} 
    &~~   & \multicolumn{2}{c}{$^{72}$Ge}  
   &~~   & \multicolumn{2}{c}{$^{136}$Xe} \\
\cline{3-4} \cline{6-7}  \cline{9-10}       
 $\mu$  &   & proton    &      neutron    &   & proton    &      neutron    & & proton    &      neutron    \\
\hline
     1 &&  3.09708E-02 &  3.76246E-02 &&  4.98732E-02 &  6.18996E-02 &&  4.64213E-02 &  6.96034E-02 \\
     2 &&  5.88325E-02 &  7.02530E-02 &&  6.19545E-02 &  7.60695E-02 &&  5.66577E-02 &  8.15343E-02 \\
     3 &&  2.10665E-02 &  2.43967E-02 && -1.96985E-02 & -1.88170E-02 && -2.89511E-02 & -3.93097E-02 \\
     4 && -2.71647E-02 & -3.07229E-02 && -3.69138E-02 & -3.41701E-02 && -4.44567E-02 & -5.02359E-02 \\
     5 && -2.58345E-02 & -2.67849E-02 &&  1.29234E-02 &  1.68135E-02 &&  2.16418E-02 &  3.33255E-02 \\
     6 &&  1.35027E-03 &  5.39599E-03 &&  2.33454E-02 &  1.51003E-02 &&  2.99702E-02 &  2.78136E-02 \\
     7 &&  1.14853E-02 &  1.73114E-02 &&  3.16112E-03 & -1.37351E-02 && -1.22571E-02 & -2.05113E-02 \\
     8 &&  3.83701E-03 &  9.32738E-03 && -8.94086E-04 & -1.58768E-02 && -2.07746E-02 & -7.97235E-03 \\
     9 && -2.49991E-03 &  2.11831E-03 &&  3.68467E-03 & -5.21935E-03 && -4.58139E-03 &  2.24455E-02 \\
    10 && -2.09091E-03 &  1.01436E-03 &&  2.56986E-03 & -1.03807E-03 && -8.95531E-04 &  1.91808E-02 \\
    11 && -1.66806E-04 &  1.16466E-03 &&  1.12422E-04 & -7.28180E-04 && -3.60592E-03 &  5.18993E-03 \\
    12 &&  3.19675E-04 &  4.98071E-04 && -2.13356E-04 & -3.48872E-04 && -1.72562E-03 &  1.49995E-03 \\
    13 &&  8.86515E-05 & -8.27115E-05 &&  7.99035E-05 &  4.73510E-06 &&  4.85598E-04 &  1.63508E-03 \\
    14 &&  - &  - &&  7.52024E-05 &  1.45816E-05 &&  4.04238E-04 &  6.51292E-04 \\
    15 &&  - &  - &&  4.16595E-05 &  1.51430E-05 &&  - &  - \\
\hline   
R [fm] && 9.0  &  9.0 && 9.0 &  9.0 && 11.0 & 11.0 \\
\hline\hline
\end{tabular}
}
\caption{\small Fourier-Bessel coefficients $A_\mu$ and radii $R$, see
  Eq. (\ref{eq:fb}), for the proton and neutron densities obtained in
  the HF + BCS calculations. 
  }
\label{tab:fb2}
\end{center} 
\end{table}

  \section*{ACKNOWLEDGMENTS} 
  G. C. thanks P. Ciafaloni and A. Incicchitti for useful discussions. 
  This work has been partially supported by the PRIN (Italy) {\sl
  Struttura e dinamica dei nuclei fuori dalla valle di stabilit\`a}, by
  the Spanish Ministerio de Ciencia e Innovaci\'on under Contract
  Nos. FPA2009-14091-C02-02 and ACI2009-1007, and by the Junta de
  Andaluc\'{\i}a (Grant No. FQM0220).

  %

\end{document}